\documentclass{ws-ijmpd}

\begin{document}

\markboth{A. Baushev}
{Dark Matter Annihilation Near a Black Hole}

%
\catchline{}{}{}{}{}
%

\title{DARK MATTER ANNIHILATION IN THE GRAVITATIONAL FIELD OF A BLACK HOLE}

\author{ANTON BAUSHEV}

\address{Bogoliubov Laboratory of Theoretical Physics, Joint Institute for Nuclear Research\\
141980 Dubna, Moscow Region, Russia\\
abaushev@gmail.com}

\maketitle

\begin{history}
\received{Day Month Year}
\revised{Day Month Year}
\comby{Managing Editor}
\end{history}

\begin{abstract}
In this paper we consider dark matter particle annihilation in the gravitational field of black
holes.  We obtain exact distribution function of the infalling dark matter particles, and compute
the resulting flux and spectra of gamma rays coming from the objects. It is shown that the dark
matter density significantly increases near a black hole. Particle collision energy becomes very
high affecting relative cross-sections of various annihilation channels. We also discuss possible
experimental consequences of these effects.

\end{abstract}


\section{Introduction.}

The Dark Matter abundance of the Universe is now estimated as $\sim 23\%$ of the total energy content\cite{cosmol}. The
most popular hypothesis about its nature is that it consists of heavy stable elementary particles surviving from the
early Universe. There are strong cosmological reasons to believe that the particles are very cold now ($\upsilon\ll
c$).

In spite of no lack of theoretical candidates for the role of dark matter particle (hereafter DMP), its nature is still
hazy. A very promising way to make the situation clearer is to observe astrophysical annihilation of the DMPs. In this
article we consider dark matter annihilation in the gravitational field of a black hole. There are several reasons why
this task is important. First of all, the greatest dark matter density in our Galaxy should occur in the center, and
then the annihilation rate there must be the highest. A possible signal from the central source is a subject of wide
speculation\cite{rewiev}. Experimental data indicate a supermassive black hole in the centre of the
Galaxy\cite{centralbh}. Secondly, the dark matter density should significantly increase near a compact massive object
(see below). Thirdly, as it will be shown in this article, some very specific physical effects can appear nearby a
black hole.

The influence of the black hole gravitational field on the dark matter annihilation has already been
considered\cite{gondolosilk99}, but the physical models were rather rough. In particular, they completely disregarded a
significant increase of the DMP velocities and strong anisotropy of their phase space distribution. Meanwhile, as it
will be shown below, the accreting DMP velocities are comparable with the speed of light. If we surmise the DMP mass to
be equal 40~{GeV}, the particle collision energy can reach hundreds of GeVs. At such high energies the cross-sections
of various annihilation channels vary significantly. This can strongly affect the annihilation rate and annihilation
products.

The plan of our work is the following: first of all we obtain the exact phase distribution function for noninteracting
particles accreting on a Schwarzschild (nonrotating) black hole in the spherically symmetric case. We assumed for
simplicity that the particles far from the black hole are nonrelativistic ($\upsilon\ll c$). This assumption is well
motivated in physically important cases, as it will be shown below. Then we calculate the number of collisions and the
product distribution, and finally compute the product outlet to the distant observer. In conclusion, we discuss the
obtained results.

\section{Dark Matter Particle Phase Distribution in the Vicinity of a Black Hole}

The gravitational field of a non-rotating black hole of mass $M_{BH}$ in vacuum can be described by the Schwarzschild
metric:
\begin{equation}
d s^2 =\left(1-\dfrac{r_g}{r}\right) c^2 d t^2 - \dfrac{d r^2}{\left(1-\dfrac{r_g}{r}\right)}- r^2 (\sin^2\!\zeta\, d \xi^2 + d \zeta^2)
\label{sch}
\end{equation}
where $r_g \equiv \dfrac{2 G M_{BH}}{c^2}$ and $c$ is the speed of light. The trajectory of a particle falling onto a
Schwarzschild black hole can be written (Ref.~\refcite{teorpol}, equations {\it 101.4} and {\it 101.5}) as:
\begin{equation}
ct = \frac{E_0}{m c^2} \int \frac{dr}{\displaystyle{\left(1-\frac{r_g}{r}\right) } \left[\left(\displaystyle{\frac{E_0}{m c^2}}\right)^2-
\left(1+\displaystyle{\frac{M^2}{m^2 c^2 r^2}}\right)\displaystyle{\left(1-\frac{r_g}{r}\right)}\right]^{1/2}}
\label{eq1}
\end{equation}

\begin{equation}
\phi =\int \frac{M dr}{r^2 \left[\displaystyle{\frac{E_0^2}{c^2}}-
\left(m^2 c^2 +\displaystyle{\frac{M^2}{r^2}}\right)\displaystyle{\left(1-\frac{r_g}{r}\right)}\right]^{1/2}}
\label{eq2}
\end{equation}
where $E_0$, $m$ and $M$ are the total energy, mass and angular momentum of the particle respectively; $E_0$ and $M$
are, of course, integrals of motion. As we have already mentioned, there are strong reasons to believe that the dark
matter particles are cold. Then their speed is mainly determined in the Galaxy by its gravitational potential and can
be estimated as a few hundred kilometers per second. Consequently, the considered particles are nonrelativistic far
from the black hole which allows us to simplify the equations. Besides, we introduce here the system of units in which
the speed of light and the gravitational radius of the black hole are equal to the dimensionless unity: $c=1$, $r_g
=1$. Then equations (\ref{eq1},\ref{eq2}) can be rewritten as:
\begin{equation}
t =\int \frac{r^2 \sqrt{r}\, dr}{(r-1)\sqrt{r^2-\alpha^2(r-1)}}
\label{eq3}
\end{equation}

\begin{equation}
\phi =\int \frac{\alpha \, dr}{\sqrt{r}\; \sqrt{r^2-\alpha^2(r-1)}}
\label{eq4}
\end{equation}
where $\alpha \equiv \frac{M}{mc}$. For the radial component $\upsilon_r$, tangential component $\upsilon_{tan}$ and the module $\upsilon$
of the particle velocity \footnote{By $\upsilon_r$, $\upsilon_{tan}$, and $\upsilon$ we imply the true, physical components of the velocity,
in contrast to the coordinate ones. For example, $\upsilon_r$ is equal to $\dfrac{\sqrt{-g_{rr}}}{\sqrt{g_{tt}}}\dfrac{d r}{d t}$ instead of $\dfrac{d r}{d t}$
} we obtain:
\begin{equation}
\upsilon_r=\dfrac{\sqrt{r^2-\alpha^2(r-1)}}{r\sqrt{r}};\qquad
\upsilon_{tan}=\dfrac{\alpha}{r}\sqrt{\dfrac{r-1}{r}}
\label{eq5}
\end{equation}
\begin{equation}
\upsilon=\dfrac{1}{\sqrt{r}}
\label{eq7}
\end{equation}
Hence, the angle $\theta$ between the direction to the center of the black hole and the particle trajectory is:
\begin{equation}
\sin\theta=\dfrac{\alpha}{r}\sqrt{r-1};\qquad
\cos\theta=\dfrac{\sqrt{r^2-\alpha^2(r-1)}}{r}
\label{eq6}
\end{equation}
From (\ref{eq7}) one can see that all the particles reaching a radius $r$ have the same velocity module. It is
convenient to introduce the quantity $\aleph$ of particles in a unit volume moving at the angle $\theta$ to the
direction to the center of the black hole in a unit solid angle. Then the quantity of the particles crossing the sphere
of the radius $r$ in a unit time in a unit solid angle in the phase space is\footnote{We use that in the Schwarzschild
metrics the surface of of a sphere of radius $r$ is equal to $4\pi r^2$.}:
\begin{equation}
4 \pi r^2 \, \aleph \, \upsilon \cos\theta \, d\tau d\Omega
\label{eq8}
\end{equation}
If $r$ is fixed, we obtain from (\ref{eq6}):
\begin{equation}
\cos\theta\, d\Omega= \cos\theta\, d(2\pi(1-\cos\theta))=\pi\, d(\sin^2 \theta)=\pi\, d\dfrac{\alpha^2(r-1)}{r^2}=\pi \dfrac{r-1}{r^2}\, d\alpha^2
\label{eq9}
\end{equation}
Besides, we substitute the interval of physical time $d\tau$ for the interval of the universal Schwarzschild time coordinate $dt=d\tau \sqrt{1-\frac{1}{r}}$.
Then we can rewrite (\ref{eq8}) as:
\begin{equation}
4 \pi^2  \,\aleph\, \dfrac{(\sqrt{r-1})^3}{r}\, d\alpha^2 dt
\label{eq10}
\end{equation}
Let us consider the particles from the interval $[\alpha ; \alpha+d\alpha]$; $\alpha$ is an integral of motion, so
$\alpha$ and $d\alpha$ remain constant along the bundle of trajectories. Since the system is stationary, the quantity
of the particles considered crossing any sphere of the radius $r$ (if they cross it at all) in a unit of the
Schwarzschild time $t$ must be equal\footnote{We use also that the Schwarzschild metrics is static.}. From (\ref{eq10})
we obtain:
\begin{equation}
4 \pi^2  \,\aleph\, \dfrac{(\sqrt{r-1})^3}{r}=\it{const}
\label{eq11}
\end{equation}
In order to find $\aleph$, we should impose a boundary condition on the infinity. From (\ref{eq11}) it follows that
$\aleph\to 0$ if $r\!\to\! \infty$, while it would be reasonable to expect that $\aleph\to n_\infty/4\pi$, where
$n_\infty$ is the dark matter particle concentration at infinity. The contradiction is a consequence of our assumption
that the velocity of the particles at infinity $\upsilon_\infty$ is zero. According to Ref.~\refcite{teorpol}, the
gravitational capture cross-section  of nonrelativistic particles is:
\begin{equation}
\sigma=4\pi r^2_g \left(\frac{c}{\upsilon_\infty}\right)^2
\nonumber
\end{equation}
The cross-section becomes infinite if $\upsilon_\infty\to 0$. In order to avoid this difficulty, we should consider
that in practical situations $\upsilon_\infty$ is never equal to $0$ exactly: for example, near the Earth the dark
matter particle speed is $\upsilon_\infty\sim 300\: km\!/\! s$. We cut the solution (\ref{eq11}) on the radius
$r_\infty$, where the velocity given by (\ref{eq7}) becomes equal to $\upsilon_\infty$. According to (\ref{eq7}),
$r_\infty=\dfrac{1}{\upsilon^2_\infty}$. The dark matter distribution at larger distances will be considered
undisturbed by the gravitational field of the black hole\footnote{This supposition is true if $\upsilon_\infty\ll c$.
In fact, if $\upsilon_\infty = 300\: km\!/\! s$, then $r_\infty=10^6 r_g$ i.e. gravitational field on the radius
$r_\infty$ is weak}. Then the quantity of the particles crossing the sphere of the radius $r_\infty$ in a unit time in
a unit solid angle in the phase space is:
\begin{equation}
4 \pi r^2_\infty \, n_\infty  \upsilon_\infty \cos\theta \: \dfrac{d\Omega}{4\pi}\:  d\tau=
r^2_\infty \, n_\infty  \upsilon_\infty \cos\theta \: d\Omega\,  d\tau
\label{eq12}
\end{equation}
According to (\ref{eq9}) $\cos\theta\, d\Omega=\pi \dfrac{r-1}{r^2}d\alpha^2$. As $r_\infty\gg 1$, we obtain
$\cos\theta\, d\Omega= \dfrac{\pi d\alpha^2}{r_\infty}$; $d\tau=dt$. Then we can rewrite (\ref{eq12}) as:
\begin{equation}
\pi \dfrac{n_\infty}{\upsilon_\infty} \, d\alpha^2 dt
\label{eq13}
\end{equation}
Comparing equations (\ref{eq11}), (\ref{eq12}) and (\ref{eq13}), we derive that the constant in (\ref{eq11}) is equal
to:
\begin{equation}
\it{const}=\pi \dfrac{n_\infty}{\upsilon_\infty}
\nonumber
\end{equation}
Substituting it into (\ref{eq11}), we obtain:
\begin{equation}
4 \pi^2  \,\aleph\, \dfrac{(\sqrt{r-1})^3}{r}=\pi \dfrac{n_\infty}{\upsilon_\infty}
\nonumber
\end{equation}
\begin{equation}
\aleph= \dfrac{n_\infty}{4 \pi \upsilon_\infty}\;\dfrac{r}{(\sqrt{r-1})^3}
\label{eq14}
\end{equation}
Deducing the initial formula (\ref{eq11}) and considering the particles crossing the sphere $r$ at the angle $\theta$
to the direction to the center, we imply that their world lines begin at infinity. It is true for the angles
\begin{equation}
\cos\theta\ge \dfrac{\sqrt{r^2-4(r-1)}}{r}\; \Xi(2-r)
\label{eq15}
\end{equation}
where
\begin{equation}
\Xi(x)=\begin{cases}
1,& x>0\\
0,& x=0\\
-1,& x<0\\
\end{cases}
\nonumber
\end{equation}
(see the details about particle trajectories in the Schwarzschild field in Ref.~\refcite{teorpol}). The world lines,
for which the condition (\ref{eq15}) fails, should begin on the event horizon. Of course, there are no particles in
these angles. Finally, we obtain for $\aleph$:
\begin{equation}
\aleph=\begin{cases}
\dfrac{n_\infty}{4 \pi \upsilon_\infty}\;\dfrac{r}{(\sqrt{r-1})^3},& \cos\theta\ge \dfrac{\sqrt{r^2-4(r-1)}}{r}\; \Xi(2-r)\\
0,& \cos\theta< \dfrac{\sqrt{r^2-4(r-1)}}{r}\; \Xi(2-r)\\
\end{cases}
\label{eq16}
\end{equation}
This is the formula which describes the collapsing particle distribution around the black hole.

\section{Calculations and Results}

Using the distribution function (\ref{eq16}) of DMPs, one can calculate the annihilation signal (measuring by a distant
observer) for various annihilation channels and various dependencies $\sigma(E)$. As an illustration, we consider the
following important model process: a particle and an antiparticle of the dark matter annihilate into two photons($\chi
\tilde \chi  \to 2 \gamma $). The DMP mass is assumed to be $m_{\mbox{\scriptsize DM}}=40$~{GeV}. The cross-section is
described by the Breit–-Wigner formula:
\begin{equation}
\sigma(E)=\sigma_0 \: \frac{(\Gamma/2)^2}{(E-E_r)^2+(\Gamma/2)^2}
\label{eq19}
\end{equation}
Here $E$ is energy in the c.m. system, $E_r$ is the resonance energy. We take $E_r=91$~{GeV}, $\Gamma=2.5$~{GeV} which
approximately corresponds to the values of $Z_0$ boson, though we do not propose any concrete annihilation process,
considering just a model problem. We suppose $\sigma_0=10^{-4}$~{$\mbox{GeV}^{-2}$} in order to provide a
characteristic weak cross-section out of the resonance $\sigma(50\mbox{GeV})\simeq 10^{-8}$~{$\mbox{GeV}^{-2}$}. The
photons were assumed to outcome spherically symmetrically in the c.m. system. The photon energy, of course, increases
with the collision energy. Besides, the centre of mass of the colliding particles is generally speaking not at rest. So
in the laboratory reference system the photons are created asymmetrically, and their energy depends on the angle of
producing.

We use stationarity of the problem to calculate the photon outgoing to infinity. The general theory of relativity
effects are taken into proper account. First, the photons undergo redshift: a photon produced on the radius $r$ with
the energy $E_0$ outcomes to a distant observer with energy
\begin{equation}
E=E_0 \: \sqrt{1-\dfrac{r_g}{r}}
\label{eq20}
\end{equation}
Second, the photons produced at the angles
\begin{equation}
\cos\theta\ge \sqrt{1-\dfrac{27}{4}\dfrac{(r-1)}{r^3}}\; \Xi(r-\frac32)
\label{eq21}
\end{equation}
are captured by the black hole\cite{podurets}.

\begin{figure}[pb]
\centerline{\psfig{file=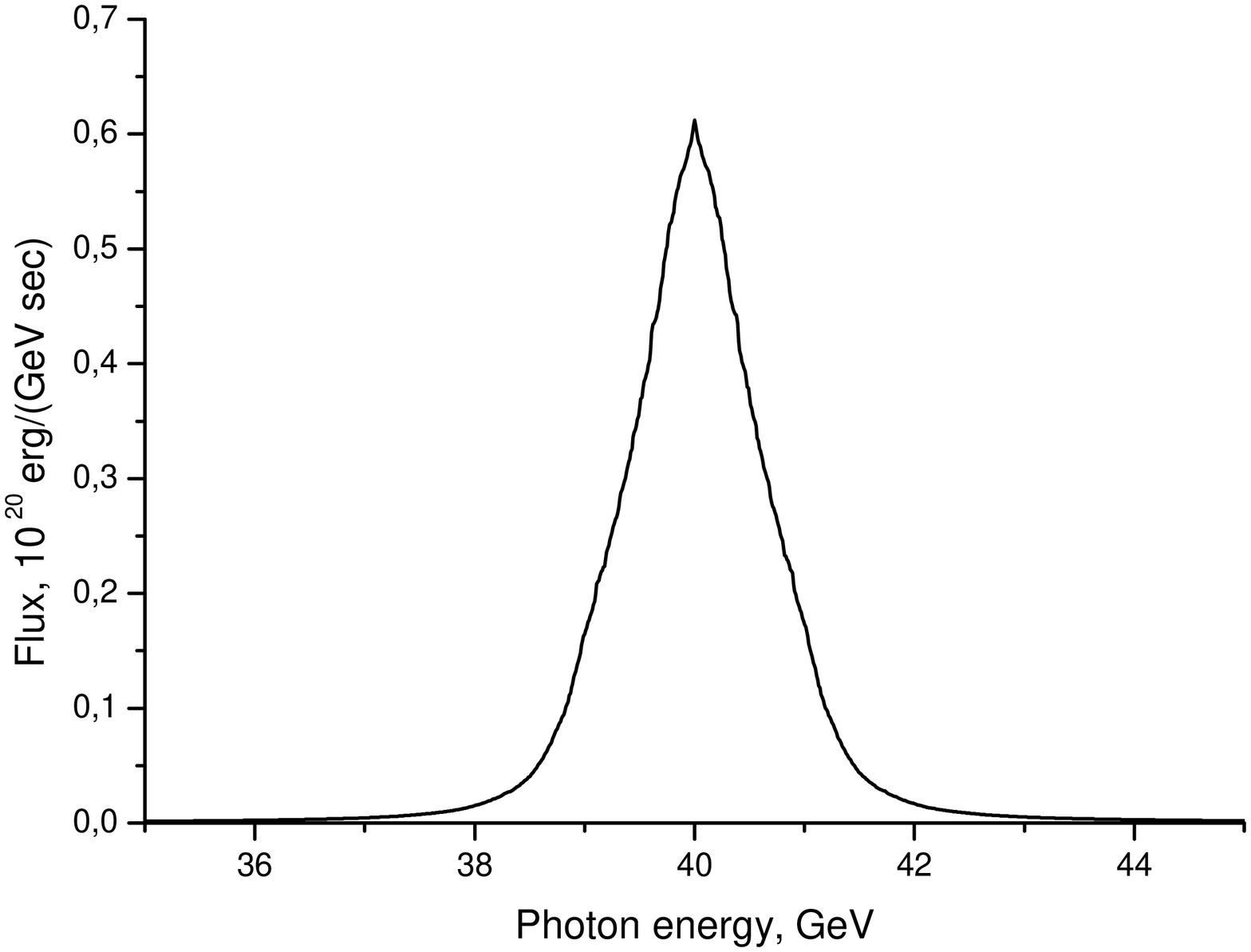,width=9.7cm,angle=270}} \vspace*{8pt} \caption{The spectrum of gamma-ray emission
produced by the dark matter annihilation into two photons. The cross-section is discribed by (\ref{eq19})} \label{pic1}
\end{figure}

We assume the black hole mass to be equal $M_{BH}=3\cdot 10^6 M_\odot$ (which corresponds to $r_g=10^{12}$~{cm}),
$\upsilon_\infty = 300$~{km/s}, and $n_\infty = 2\cdot 10^{-2}$~{$\mbox{cm}^{-3}$} ($\sim
0.8$~$\mbox{GeV}/\mbox{cm}^3$). The last value is characteristic at the Sun system vicinity, for the Galaxy centre it
must be underestimated. Unfortunately, there is no reliable estimation for the dark matter density in the centre of the
Galaxy. The results of the calculations are represented in Fig.~\ref{pic1}.

\section{Discussion}

Let us discuss the obtained results, giving special heed to the possible observable effects.

First of all, according to (\ref{eq16}), particle concentration near the black hole ($r\sim r_g$) is proportional to
$n_\infty /\upsilon_\infty$ (we should remind that $n_\infty$ and $\upsilon_\infty$ are concentration and velocity of
DMPs undisturbed by the gravitational field). Since the annihilation signal intensity is proportional to the squared
concentration, the signal grows by the factor of the order of $(c/\upsilon_\infty)^2$ with respect to the undisturbed
dark matter field. As it has been already mentioned, the dark matter is cold; the speed of its thermal motion is
negligible. The DMP velocity in the Galaxy is mainly determined by the gravitational potential of the sighting point.
In the vicinity of the Sun system $\upsilon_\infty\simeq 300$~{km/s}. Near the Galaxy centre $\upsilon_\infty$ is
bigger but of the same order. Consequently, one can expect the signal augmentation by six orders only due to the dark
matter concentration growth. Unfortunately, the effective region volume is not so large (of an order of $\pi r^3_g$),
but for the supermassive central black hole in our Galaxy $r_g\simeq 10^7$~{km}.

Secondly, though the DMP speed near the black hole tends to the speed of light, the energy of their collisions does not
excel a finite limit. Indeed, using (\ref{eq5},\ref{eq7}) one can make sure that the sum energy of two colliding DMPs
in the center-of-mass system is equal:
\begin{equation}
E=2 m \dfrac{\sqrt{1-\upsilon^2 \cos^2 \beta}}{\sqrt{1-\upsilon^2}}=2 m \dfrac{\sqrt{r-\cos^2 \beta}}{\sqrt{r-1}}
\label{eq17}
\end{equation}
where $2\beta$ is the angle between the velocities of the particles. The last radius, where the head-on collisions
(with $\beta=\pi/2$) are still possible, is $r=2$. The corresponding collision energy is $2\sqrt{2}m$. When particles
get closer to the horizon, their speeds remain to grow, but possible $\beta$ rapidly decreases (since the velocities of
all the falling particles become almost parallel). When $r<2$, it is determined by (\ref{eq15}). Substituting
(\ref{eq15}) to (\ref{eq17}), we obtain:
$$
E_{max}=2 m \sqrt{r^2+4}\nonumber
$$
\begin{equation}
\lim_{r\to 1} E_{max}= 2\sqrt{5}\; m
\label{eq18}
\end{equation}
Thus, the maximal collision energy nowhere surpasses $2\sqrt{5}\; m$. Practical conclusion from this fact is that if
the threshold energy of some DMP interaction process excels $2\sqrt{5}\; m$, the process is impossible. Nevertheless,
for 40~{GeV} particles the limit is $\sim 180$~{GeV}.

So the energy of the DMP collisions can be very high, which can significantly affect the cross-sections of various
annihilation channels and even change the main one. In particular, strong resonances at the energies around $Z_0$ and
Higgs bosons masses can appear. Of course, it could change the annihilation products and strongly influence the
annihilation rate.

Thirdly, let us consider the resulting spectrum of the annihilation signal. Under usual situations (far from compact
objects) the considered way of annihilation ($\chi \tilde \chi  \to 2 \gamma $) should produce a very narrow line: the
width of photon spectral distribution is determined by the particle energy spread:
\begin{equation}
\Delta E\sim m_{\mbox{\scriptsize DM}}\upsilon^2
\label{eq22}
\end{equation}
In the considered case there are at least three mechanisms of the line broadening and shifting. First, the annihilating
particle collision energy significantly depends on $r$ and the angle between particle velocities (see
equation~\ref{eq17}). The energy of produced photons varies correspondingly. Second, the centre of mass of the
colliding particles is generally speaking not at rest. Consequently, Lorentz shifting appears. Third, gravitational
redshift quite strongly decreases the energy of the photons generated near $r_g$. We should mention that the effects of
the first and third mechanisms are opposite in sign: so they can compensate each other. Nevertheless, one could expect
that the annihilation line was arbitrarily wide.

In Fig.~\ref{pic1} we can see that various broadening effects well compensate each other, and the line remains
respectively narrow and unshifted. This fact is of big practical significance. Indeed, different ways of dark matter
annihilation signal detection from the central Galactic source (where a supermassive black hole is situated, as we have
already mentioned) are discussed in the literature\cite{rewiev}. It is suggested to observe immediate annihilation
products as well as the synchrotron emission radiated by them. However, a serious problem appears here. The
concentration of not only dark, but of normal matter too is high in the centre of the Galaxy. The normal matter
accretion on the black hole is accompanied by complex magnetohydrodynamic effects which can generate high-energy
particles, synchrotron emission etc. of various spectra and characteristics. Since the cross-section of interaction of
baryonic matter is incommensurably greater, effectiveness of these processes is very big. Therefore, even if some
interesting signal from the Galactic centre has been observed, one should prove that it is the dark matter annihilation
that gives it. The case when the emission spectrum is wide it's a very difficult task. Whatever is the signal, to fit
it by the ordinary matter emission is usually possible. In the case of a narrow line the situation is much better. It
is more complicated to offer an alternative explanation to a narrow line on the energy $>10$~{GeV}. Moreover, the
energy of the line would immediately give us the DMP mass. Therefore, the study of the narrow annihilation line
generation opportunity is a very promising problem, even if the relative cross-section of the channel is small. As we
have seen, the DMP density near a black hole strongly increases. Besides, the collision energy growth can strongly
elevate the two-photon annihilation channel efficiency. Thus, we can expect an intensive line formation about a black
hole. A misgiving appears, however, that the line can be transformed into a wide band ($\Delta E/E\sim 1$) by the
gravitational redshift, for instance. Figure~\ref{pic1} shows that it does not happen. The line remains narrow enough
($\Delta E/E\sim 0.05$). On the other hand, the line turns out to be distinctly wider then if it appears far from
compact objects. In the last case, the width can be estimated via (\ref{eq22}) which gives us $\Delta E/E\sim
(\upsilon_\infty / c)^2\simeq 10^{-6}$. So the line generated near a black hole can be easily distinguished from the
line appearing under usual situations.

\end{document}